\documentclass[%
 aip,
 cp,  
 amsmath,
 amssymb,
 reprint,%
]{revtex4-2}

\usepackage{graphicx}
\usepackage[caption=false]{subfig}  
\usepackage{dcolumn}
\usepackage{bm}
\usepackage{color}

\usepackage[utf8]{inputenc}
\usepackage[T1]{fontenc}
\usepackage{mathptmx}

\usepackage{fancyhdr,lastpage}
\usepackage{hyperref}
\hypersetup{
    colorlinks=true,
    linkcolor=blue,
    citecolor=blue,     
    urlcolor=blue,
    breaklinks=true
}

\DeclareMathOperator{\Imag}{Im}
\DeclareMathOperator{\Real}{Re}
\DeclareMathOperator{\sech}{sech}

\begin{document}

\title{Twenty-Five Years of Dissipative Solitons}

\author{Ivan C.\ Christov} 
\email[Corresponding author: ]{christov@purdue.edu}
\homepage{http://tmnt-lab.org}

\author{Zongxin Yu}
\email{yu754@purdue.edu}
 
\affiliation{School of Mechanical Engineering, Purdue University, West Lafayette, Indiana 47907, USA}


\begin{abstract}
In 1995, C.~I.~Christov and M.~G.~Velarde introduced the concept of a dissipative soliton in a long-wave thin-film equation  [\textit{Physica D} \textbf{86}, 323--347]. In the 25 years since, the subject has blossomed to include many related phenomena. The focus of this short note is to survey the conceptual influence of the concept of a ``production-dissipation (input-output) energy balance'' that they identified. Our recent results on nonlinear periodic waves as dissipative solitons (in a model equation for a ferrofluid interface in a parallel-flow rectangular geometry subject to an inhomogeneous magnetic field) have shown that the classical concept also applies to nonlocalized (specifically,  spatially periodic) nonlinear coherent structures. Thus, we revisit the so-called KdV-KSV equation studied by C.~I.~Christov and M.~G.~Velarde to demonstrate that it also possesses spatially periodic dissipative soliton solutions. These coherent structures arise when the linearly unstable flat film state evolves to sufficiently large amplitude. The linear instability is then arrested when the nonlinearity saturates, leading to permanent traveling waves. Although the two model equations considered in this short note feature the same prototypical linear long-wave instability mechanism, along with similar linear dispersion, their nonlinearities are fundamentally different. These nonlinear terms set the shape and eventual dynamics of the nonlinear periodic waves. Intriguingly, the nonintegrable equations discussed in this note also exhibit multiperiodic nonlinear wave solutions, akin to the polycnoidal waves discussed by J.~P.~Boyd in the context of the completely integrable KdV equation.
\end{abstract}

\maketitle

\section{Introduction}

According to C.~I.~Christov and M.~G.~Velarde \cite{CV95}, \emph{dissipative solitons} are localized solutions of nonconservative nonlinear evolution (or wave) equations. To illustrate the concept, consider the original example of the (suitably nondimensionalized) Korteweg--de Vries--Kuramoto--Sivashinsky--Velarde (KdV-KSV) partial differential equation (PDE):
\begin{equation}
    \underbrace{\eta_t + 2a_1 \eta\eta_x + a_3 \eta_{xxx}}_{\text{KdV}} + \underbrace{a_2 \eta_{xx} + a_4 \eta_{xxxx}}_{\text{KS}} + \underbrace{a_5 (\eta \eta_x)_x}_{\text{Velarde}} = 0,\qquad x\in\Omega, \quad t>0,
    \label{eq:KdV-KSV}
\end{equation}
where the diffusion coefficients are positive, $a_{2,4}>0$, and $\Omega\subset\mathbb{R}$. In equation~\eqref{eq:KdV-KSV}, $\eta=\eta(x,t)$ generically represents the dimensionless elevation of a thin film's free surface. The KdV terms in equation~\eqref{eq:KdV-KSV} are perturbed by the two prototypical terms from the KS equation, and an extra term due to Velarde and co-workers (see, e.g., \cite{GV91,GV92}) captures the Marangoni effect (flow driven by surface tension gradients). More generally, Eq.~\eqref{eq:KdV-KSV} represents a \emph{model equation} for wavy viscous flow, capturing the leading-order effects on the evolution of the film's free surface \cite{H74} (see also \cite{KRSV12} for discussion and background on the gravity-driven case).

By the so-called energy method \cite{St04}, one can form a ``budget'' for $\mathcal{E}(t) := \tfrac{1}{2} \int_\Omega \eta(x,t)^2 \,\mathrm{d}x$ by multiplying  Eq.~\eqref{eq:KdV-KSV} by $\eta(x,t)$ and integrating over $\Omega$:
\begin{equation}
    \frac{\mathrm{d}\mathcal{E}}{\mathrm{d}t} =  \underbrace{a_2 \int_\Omega \eta_x^2 \,\mathrm{d}x}_{\text{production}} + a_5 \int_\Omega \eta\eta_x^2 \,\mathrm{d}x - \underbrace{a_4 \int_\Omega \eta_{xx}^2 \,\mathrm{d}x}_{\text{dissipation}}.
    \label{eq:balance}
\end{equation}
Equation~\eqref{eq:balance} is, in fact, a balance law describing the competition of \emph{production} and \emph{dissipation} (recalling the positive signs of $a_{2,4}>0$ were fixed above). Importantly, the middle term in the balance in Eq.~\eqref{eq:balance} has an \emph{indefinite} sign. In other words, this term can be positive or negative (i.e., be production or dissipation) depending on the spatiotemporal evolution of $\eta(x,t)$. Further, it follows that Eq.~\eqref{eq:KdV-KSV} is \emph{nonconservative} because $\mathrm{d}\mathcal{E}/\mathrm{d}t$ does not vanish identically $\forall \eta$.

The domain of integration has not been explicitly specified to allow for the consideration of two cases in which the boundary contributions (after integration by parts) vanish: (i) $
\Omega = \mathbb{R}$ and asymptotic boundary conditions ($\eta,\eta_x,\eta_{xx},\hdots\to0$ as $|x|\to\infty$), and (ii) $\Omega = [0,2\pi]$ and periodic boundary conditions ($\eta,\eta_x,\eta_{xx},\hdots|_{x=0}^{x=2\pi}=0$). Only the first case was considered in the original paper \cite{CV95}. In this short note, we consider the second case.

Although originally framed in the context of the nonlinear \emph{evolution} equation~\eqref{eq:KdV-KSV}, the dissipative soliton concept applies equally well to nonlinear \emph{wave} equations, which support bidirectional propagation \cite{CV94,C02,KMJC11}. In a paper in the same special issue as \cite{CV95}, M.~Bode and H.-G.~Purwins \cite{BP95} discussed dissipative solitons in the context of pattern formation in reaction-diffusion systems. This latter topic has become well developed over the years \cite{AA05}, but it is beyond the scope of the present discussion focusing on model nonconservative equations of nonlinear dispersive waves.

C.~I.~Christov and M.~G.~Velarde \cite{CV95}, considered two situations. In the first case, $0<a_{2,4,5}\ll1$, and the nonconservative terms are a small perturbation to KdV. In this case, ``Zabusky and Kruskal's soliton concept is extended in two directions: [first] to ``long'' transients practically ``permanent'' and solitonic''. Specifically, taking $a_{2,4,5}=\mathrm{O}(\varepsilon)$, the so-called $\sech^2$ soliton solution of KdV was shown to persist in KdV-KSV for long times, up to $t=\mathrm{O}(\varepsilon^{-1})$. Second, the more interesting case is when the right-hand side of Eq.~\eqref{eq:balance} vanishes for $a_{2,4,5}\ne0$, in which case ``true permanent wave-particles with, however, inelastic behaviour upon collisions'' were discovered. This latter case corresponds to the dissipative solitons. Although the mathematical literature strictly defines a soliton as the exact solitary wave solution of an \emph{integrable} equation, like KdV, here we take the physicist's point of view that if the solitary wave has some generic interaction property, then we shall call it a soliton to not belabor the point (see, e.g., \cite{S04}, p.~849).

Motivated by some earlier studies demonstrating nonlinear waves on ferrofluid interfaces \cite{LM12,YC21}, we recently introduced \cite{YC21b} a new long-wave equation for a ferrofluid interface subject to an angled nonuniform external magnetic field. This equation is of the \emph{generalized} (i.e., dispersive) KS type, and takes the form (suitably nondimensionalized):
\begin{equation}
    \underbrace{\eta_t + \delta\alpha \eta_{xx} + \eta_{xxxx}}_{\text{linear KS}} 
    - \underbrace{\beta \eta_{xxx}}_{\text{dispersion}} + \underbrace{[(\delta\alpha \eta_x - \beta \eta_{xx} + \eta_{xxx} )\eta]_x - \delta(\gamma \eta_x^2 )_{xx}}_{\text{nonlinearity}} = 0,\qquad x\in\Omega,\quad t>0,
    \label{eq:ferro_KS}
\end{equation}
where $\delta\alpha>0$. Physically, the constants $\alpha$, $\beta$ and $\gamma$ are determined by the magnetic Bond numbers, which express the ratio of magnetic field strength (in the $x$- or $y$-direction) to surface tension \cite{YC21b}. Meanwhile, $\delta\ll1$ is a geometric parameter expressing the long-wave approximation. For the present purposes, $\alpha$, $\beta$, $\gamma$ and $\delta$ in Eq.~\eqref{eq:ferro_KS} can be taken to be generic constants, like $a_{1,2,3,4,5}$ in Eq.~\eqref{eq:KdV-KSV}.

The nonlinearity in Eq.~\eqref{eq:ferro_KS} is now quite complex, compared to the KdV and Velarde terms (i.e., $2a_1\eta\eta_x$ and $a_5(\eta\eta_x)_x$) in Eq.~\eqref{eq:KdV-KSV}. The added complexity is due to the force balance between surface tension and magnetic tractions on the ferrofluid interface \cite{YC21b}. 

The energy budget for Eq.~\eqref{eq:ferro_KS} is
\begin{equation}
    \frac{\mathrm{d}\mathcal{E}}{\mathrm{d}t} = \underbrace{\delta\alpha \int_\Omega \eta_x^2 \,\mathrm{d}x}_{\text{production}} + \delta\alpha \int_\Omega \eta\eta_{x}^2 \,\mathrm{d}x + \frac{1}{2}\beta \int_\Omega \eta_{x}^3  \,\mathrm{d}x \overbrace{- \int_\Omega \eta \eta_{xx}^2 \,\mathrm{d}x  - \underbrace{\int_\Omega \eta_{xx}^2 \,\mathrm{d}x}_{\text{dissipation}}}^{\text{due to surface tension}}.
    \label{eq:balance2}
\end{equation}
Now, three terms on the right-hand side of the energy budget remain sign-indefinite \textit{a priori}. Nevertheless, observe that the sign-indefinite terms in Eq.~\eqref{eq:balance2} are related to the complex nonlinearity arising from the interfacial force balance on the fluid, just like the sign-indefinite term in Eq.~\eqref{eq:balance} comes from the Velarde term that accounts for the Marangoni effect at the interface. The sign-definite production and dissipation terms in Eqs.~\eqref{eq:balance} and \eqref{eq:balance2} are the same, arising from the linear KS equation's energy production and dissipation mechanisms.

\section{Nonlinear periodic waves as dissipative solitons}

Previously, localized (solitary) waves and kinks (topological solitons) were considered as dissipative solitons \cite{CV95,KMJC11}. Here, we highlight the possibility of nonlinear periodic waves as dissipative solitons. That is, these dissipative solitons are spatially periodic, rather than spatially localized. The mechanism underlying the generation of these nonlinear periodic waves is the long-wave linear instability of the flat base state ($\eta=0$), which is ultimately ``arrested'' by the saturation of the nonlinearity \cite{BP98}. 

\subsection{Energy phase plane concept}

A useful concept for studying the evolutionary dynamics of dissipative solitons out of some initial conditions (see, e.g., Eq.~\eqref{eq:perturbed_IC} below) is the energy phase plane $(\mathcal{E},\dot{\mathcal{E}})$ employed in \cite{KKP15,YC21b}. For convenience, we introduced the over-dot notation $\dot{\mathcal{E}}\equiv\mathrm{d}\mathcal{E}/\mathrm{d}t$. During the evolution from the initial condition, $\eta(x,t)$ will be such that $\mathcal{E}(0),\dot{\mathcal{E}}(0)\ne0$. If the evolution leads to a dissipative soliton, we expect that $(\dot{\mathcal{E}}, \mathcal{E})\to (0, \mathcal{E}^*)$, as $t\to t^*$, where $t^*$ is some transient timescale over which the dissipative solution emerges. Then, $\mathcal{E}^* \equiv \mathcal{E}(t^*)$ is the dissipative soliton's finite energy (conserved in the absence of further perturbations).

\subsection{KdV-KSV equation}

To determine the instability of the flat base state $\eta=0$ under  Eq.~\eqref{eq:KdV-KSV}, we expand the interface shape as $\eta(x,t) = 0 + \epsilon e^{i(kx-\omega t)} + c.c.$ ($\epsilon\ll1$ being an arbitrary perturbation strength), where `$c.c.$' denotes `complex conjugate,' and $i=\sqrt{-1}$. Then, we substitute this form for $\eta$ into Eq.~\eqref{eq:KdV-KSV} and neglect terms of $\mathrm{O}(\epsilon)$. This calculation yields the \emph{dispersion relation} between the perturbation time-frequency $\omega$ and the perturbation wavenumber $k$:
\begin{equation}
    \omega(k) = - a_3 k^3 + i a_2 k^2 - i a_4 k^4.
    \label{eq:disp_rel_KdV-KSV}
\end{equation}
Observe that $\omega(k)\in\mathbb{C}$, i.e., the dispersion relation is complex, thus waves can propagate. Specifically, linear waves under Eq.~\eqref{eq:KdV-KSV} are dispersive with phase velocity $v_p(k) = \Real[\omega(k)]/k=-a_3 k^2$.

Now, let $k_c$ solve $\Imag[\omega(k)]=0$; without loss of generality, we keep $k_c>0$. If $k>k_c = \sqrt{a_2/a_4}$, then the waves are damped, and the perturbation returns back to the flat state ($\eta(x,t)\to0$ as $t\to\infty$). Meanwhile, for $k<k_c$, the perturbation grows exponentially in time (linear instability). This type of unstable band and quartic structure of $\Imag[\omega(k)]$ (competition between diffusion and ``anti-diffusion'') is typical of pattern-forming thin-film systems \cite{ODB97}. The unstable case of $k<k_c$ is the one of interest henceforth.

\begin{figure}[t]
\includegraphics[width=\textwidth]{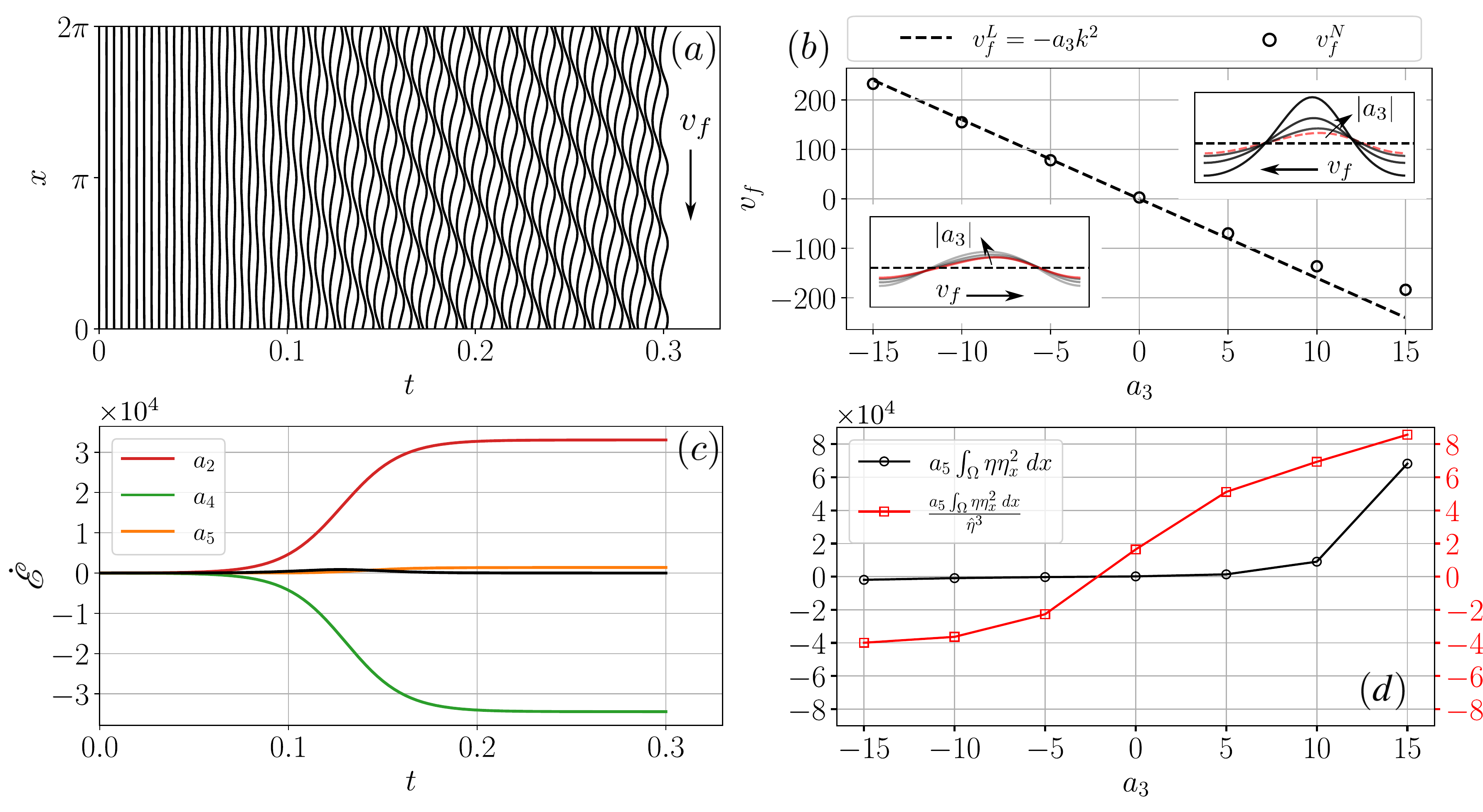}
\caption{(a) Space-time plot of the nonlinear evolution of the interface into a permanent traveling wave, under Eq.~\eqref{eq:KdV-KSV}, starting from a small perturbation of the flat base state [i.e., taking $\eta(x,0)=0.1\cos (4x)$]. The coefficients of the PDE are taken to be $a_1=18$, $a_2=18$, $a_4=1$, $a_5=1$, and $a_3=5$. (b) The dependence of the propagation velocity $v_f$ ($v_f^L=v_p$ from the linear dispersion relation and $v_f^N$ extracted from the PDE simulation) and corresponding wave profiles' shape on the dispersion parameter $a_3$ (the red curves in the insets are for $a_3=0$), with other $a_i$ same as (a). (c) Energy budget of the nonlinear traveling wave generation process shown in (a). The black curve shows the sum of the components in Eq.~\eqref{eq:balance}, which is seen to approach zero as the wave evolves into a dissipative soliton. (d) Variation of the energy budget component involving $a_5$ in Eq.~\eqref{eq:balance}, from dissipation to production, as the dispersion coefficient $a_3$ is changed. The black curve (circles) shows the absolute contribution, while the red curve (squares) shows the normalized one with respect to  $\hat{\eta}^3$, where $\hat{\eta}=\max_{0\le x\le 2\pi}|\eta(x,t^*)|$ is the amplitude of the corresponding traveling wave profile.}
\label{fig:1}
\end{figure}

To understand how saturation of the nonlinearity arrests the exponential linear instability, leading to a permanent traveling wave, we solve Eq.~\eqref{eq:KdV-KSV} on $\Omega=[0,2\pi]$ subject to periodic boundary conditions $\eta(x,t) = \eta(x+2\pi,t)$ $\forall t\ge0$, starting from the initial condition
\begin{equation}
    \eta(x,0) = 0 + \epsilon \cos(k_0 x), \qquad k_0 \le k_c, \qquad \epsilon\ll 1.
    \label{eq:perturbed_IC}
\end{equation}
The numerical method employed is a Fourier pseudospectral method \cite{Boyd00} for the spatial derivatives with an exponential time-differencing fourth-order Runge--Kutta (ETDRK4) \cite{KT05}; see \cite{YC21b} for details on its benchmarking. Observe that, for this initial condition, we would have $\mathcal{E}(0) = \mathrm{O}(\epsilon^2)$.

Figure~\ref{fig:1}(a) shows one example for which the KdV-KSV equation exhibits a nonlinear periodic traveling wave solution. With the given parameters, the critical wave number is $k_c\approx 4.24$, such that the initial wave number $k_0=4$ is subject to a weak linear instability. Figure~\ref{fig:1}(c) shows the energy budget, including contributions from the production term (multiplied by $a_2$), the dissipation term (multiplied by $a_4$), and the indefinite term (multiplied by $a_5$) in Eq.~\eqref{eq:balance}. Upon achieving a balance between these three terms, the nonlinear traveling wave solution emerges. In other words, the dynamics approaches an equilibrium point in the energy phase plane and the energy change rate $\dot{\mathcal{E}}\rightarrow 0$.

Figure~\ref{fig:1}(b) shows the dependence of the propagation velocity $v_f$ on the dispersion parameter $a_3$. The linear phase velocity can effectively predict the nonlinear propagation velocity of the wave, i.e., $v_f\approx v_p(k=4)$, for small-amplitude solutions in the range $a_3\in [-15,5]$. Note that, for $a_3=0$, $v_f\neq 0$ due to the contribution from the nonlinear advective term, $2a_1\eta\eta_x$, in Eq.~\eqref{eq:KdV-KSV}. This term, termed a Hopf nonlinearity, controls the dependence between amplitude and propagation velocity, especially for left-propagating waves ($a_3>0$).

Figure~\ref{fig:1}(d) shows the distinct role of the $a_5$ (Velarde) term in the energy budget. As can be inferred from the plot, this term can be \emph{either} production or dissipation. For right-propagating waves ($a_3<0$), $a_5\int_\Omega \eta\eta_x^2 dx<0$, and this term dissipates energy in the system. For left-propagating waves ($a_3>0$), $a_5\int_\Omega \eta\eta_x^2 dx>0$, and this term serves as energy production. This contribution becomes more important as the wave amplitude increases (with $a_3$).

\subsection{Long-wave equation for a ferrofluid thin film}

As before, it can be shown that the stability of the flat base state $\eta=0$ under Eq.~\eqref{eq:ferro_KS} is characterized by the dispersion relation:
\begin{equation}
    \omega(k) = \beta k^3 + i \delta \alpha k^2 - i k^4.
    \label{eq:disp_rel_ferro_KS}
\end{equation}
Now, $k_c = \sqrt{\delta\alpha}$ \cite{YC21b}. Again, linear waves are dispersive with phase velocity $v_p(k) = \Real[\omega(k)]/k=\beta k^2$. Indeed, the long-wave linear instability is the same in both model equations, being set by the linear KS-type terms. The main differences between Eqs.~\eqref{eq:KdV-KSV} and \eqref{eq:ferro_KS} emerge from their nonlinear terms. Obviously, it is expected that these different nonlinearities will saturate differently, and lead to different dissipative soliton dynamics. To highlight the latter differences, as before, we solve Eq.~\eqref{eq:ferro_KS} numerically on $\Omega=[0,2\pi]$, subject to periodic boundary conditions, starting from the initial condition in Eq.~\eqref{eq:perturbed_IC}.

\begin{figure}[t]
\includegraphics[width=\textwidth]{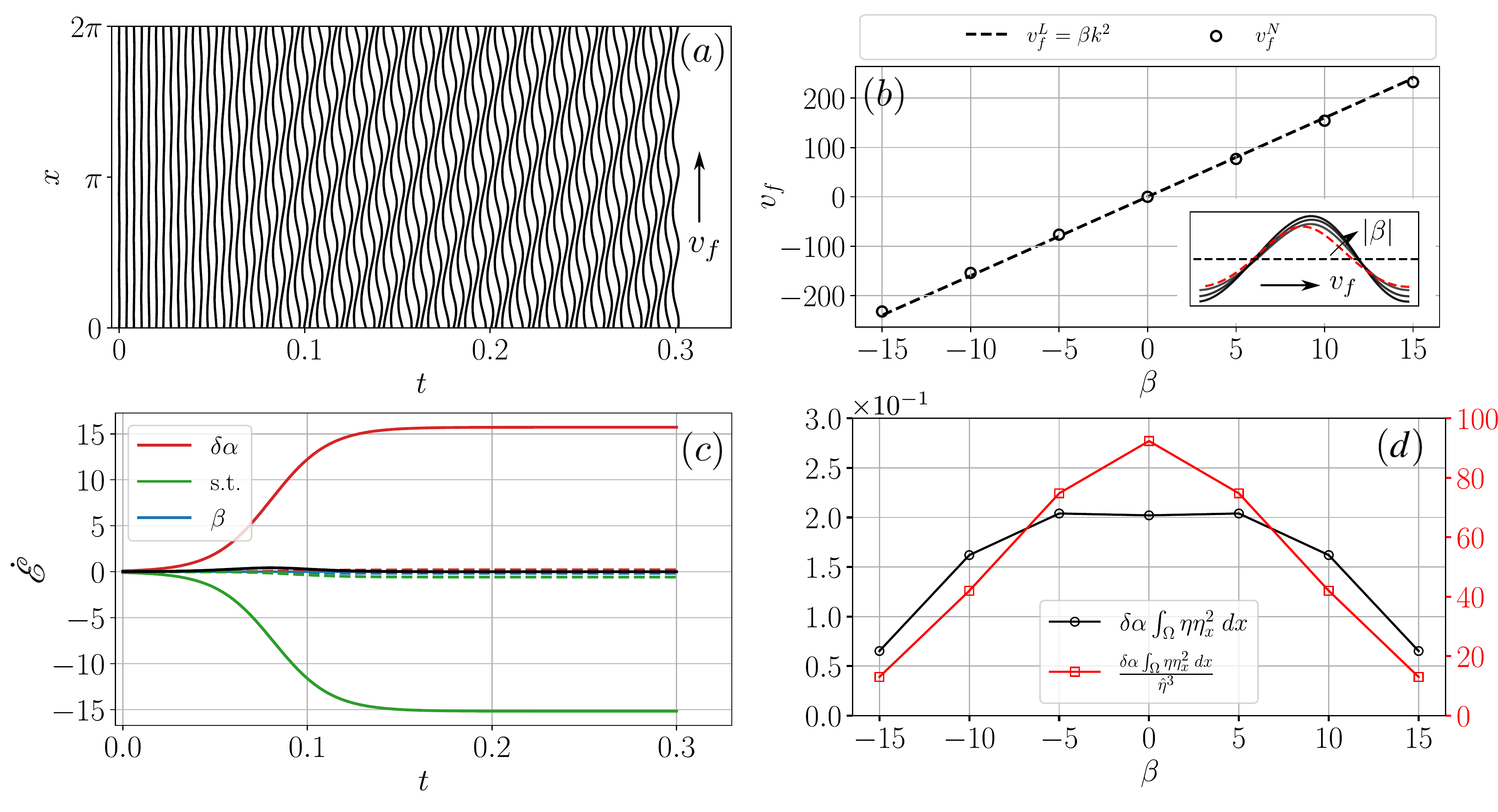}
\caption{(a) Space-time plot of the nonlinear evolution of the interface into a permanent traveling wave, under Eq.~\eqref{eq:ferro_KS}, starting from a small perturbation of the flat base state [i.e., taking $\eta(x,0)=0.01\cos (4x)$]. The coefficients of the PDE are taken to be $\delta \alpha=18$, $\beta=5$, and $\gamma=44.93$. (b) The dependence of the propagation velocity $v_f$ ($v_f^L=v_p$ from the linear dispersion relation and $v_f^N$ extracted from the PDE simulation) and corresponding wave profiles on the dispersion parameter $\beta$ (the dashed red curve in the inset is for $\beta=0$), with $\delta \alpha=18$, and $\gamma$ determined by physical relations in \cite{YC21b}. (c) Energy budget of the nonlinear traveling wave generation process shown in (a). The dashed curves represent the contributions from nonlinear terms, the black curve shows the sum of the components in Eq.~\eqref{eq:balance2}, and `s.t.' stands for `surface tension'. (d) Variation of the energy budget component involving $\delta\alpha$ in Eq.~\eqref{eq:balance2}, from dissipation to production, as the dispersion coefficient $\beta$ is changed. The black curve (circles) shows the absolute contribution, while the red curve (squares) shows the normalized one with respect to $\hat{\eta}^3$, where $\hat{\eta}=\max_{0\le x\le 2\pi}|\eta(x,t^*)|$ is the amplitude of the corresponding traveling wave profile.}
\label{fig:2}
\end{figure}

Figure~\ref{fig:2}(a) shows the emergence of the nonlinear periodic traveling wave for a critical wave number $k_c\approx 4.24$, in which case the initial perturbation is subject to weak linear instability. Meanwhile Fig.~\ref{fig:2}(c) shows the dynamics' energy budget (and its evolution). The $\delta\alpha\eta_{xx}$ term from the external magnetic field leads to energy production, while the $\eta_{xxxx}$ from surface tension represents energy dissipation in the balance in Eq.~\eqref{eq:balance2}. These two terms' contributions dominate over those from the nonlinear terms in the energy balance.

Again, the nonlinear wave's propagation velocity can be well predicted by the linear dispersion relation, namely $v_f\approx v_p(k)=\beta k^2$, for both directions of propagation. Both the velocities and the wave profiles preserve their symmetry under the transformation $\beta \rightarrow -\beta$, which is required by the physics that it represents (i.e., inverting the magnetic field direction inverts the direction of wave propagation). Note that, under Eq.~\eqref{eq:ferro_KS}, the dependence of the propagation velocity on the wave amplitude is less significant than under Eq.~\eqref{eq:KdV-KSV} (shown in Fig.~\ref{fig:1}(b)) for a similar propagation velocity range (see \cite{YC21b} for a wider parameter scope). 

Figure~\ref{fig:2}(d) shows the different role of the nonlinear $\delta \alpha$ term in the energy balance in Eq.~\eqref{eq:balance2}. The variation of this component with the dispersion parameter is different under Eq.~\eqref{eq:ferro_KS} than under the KdV-KSV equation~\eqref{eq:KdV-KSV}, which was shown in Fig.~\ref{fig:1}(d), since the energy balance in Eq.~\eqref{eq:balance2}  involves two more sign-indefinite terms. The role of the $\delta \alpha$ term as energy production becomes weaker as  $|\beta|$ increases. The plot in Fig.~\ref{fig:2}(d) is symmetric about $\beta=0$ due to the symmetry of the wave profiles under the transformation $\beta\to-\beta$, as discussed above.

\subsection{Spatially multiperiodic dissipative soliton and transition under KdV-KSV}

\begin{figure}[t]
\includegraphics[width=\textwidth]{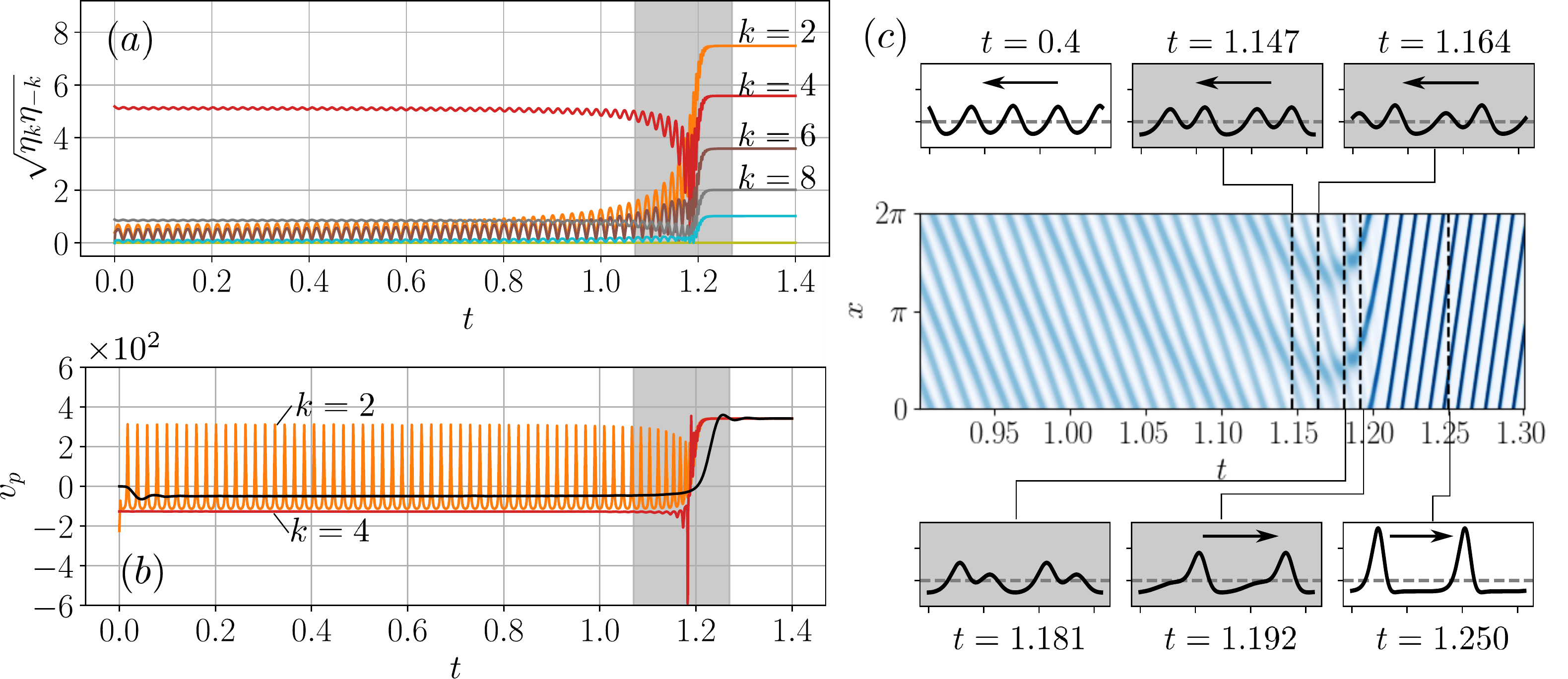}
 \caption{(a) Fourier modes' energy evolution and interactions for a perturbed period-four nonlinear traveling wave $\Theta_4(\zeta)$ of the KdV-KSV equation~\eqref{eq:KdV-KSV}. This period-two perturbation is taken to be $\eta_p(\zeta)=0.8\sin(2 \zeta)$. Here, $a_1=20$, $a_2=19.1$, $a_3=10$, $a_4=1$, and $a_5=1$. (b) The phase velocities of mode 4 and mode 2. The black solid curve shows the filtered Fourier mode phase velocity $v_p(k=2)$. (c) Space-time plot and the corresponding wave profiles during the transition period, $t\in[0.9,1.3]$.}
 \label{fig:a2a4}
\end{figure}

As shown above, perturbations of the flat state grow into stable nonlinear periodic traveling waves. Let us denote such a nonlinear solution as $\Theta_n(\zeta)$ if it has period-$n$, where $\zeta=x-v_f t$ is the traveling wave coordinate. Next, we perturb the period-four traveling wave profile $\Theta_4(\zeta)$, obtained under the parameter choices of $a_1=20$, $a_2=19.1$, $a_3=10$, $a_4=1$, $a_5=1$, by taking $\eta(x,0) = \Theta_4(\zeta) + \eta_p(\zeta)$, where $\eta_p(\zeta) = 0.8\sin(2 \zeta)$, as the initial condition for a simulation.  We track the evolution of this perturbed traveling wave via direct simulation of the PDE~\eqref{eq:KdV-KSV}. The wave profile $\eta$ is decomposed into complex Fourier modes $\{\eta_k\}$, and the evolution of their energy, $|\eta_k| = \sqrt{\eta_k\eta_{-k}}$, is shown in Fig.~\ref{fig:a2a4}(a). 

We observe a long-lived interaction between Fourier modes 2 and 4 for $t\in[0.,1.1]$. This interaction becomes stronger and strong, until a transition occurs during $t\in[0.9,1.25]$, which is also observed in the space-time plot of the wave profile's evolution in Fig.~\ref{fig:a2a4}(c). Specifically, we observe that the period-four traveling wave is modulated by mode 2, and this coexistence lasts for a relatively long time (compared to the total transition time), until mode 2 ultimately becomes dominant for $t>1.4$.

For this set of parameters, the system is considered strongly dispersive, and the KdV terms dominate in Eq.~\eqref{eq:KdV-KSV}. In this case, it is therefore expected that the traveling wave solution bears similarity to cnoidal waves (see section 5 in \cite{c12}), which can also be approximated by low-dimensional Fourier series. Such numerically identified long-lived multiperiodic wave states can be thought of as analogues to polycnoidal waves (multiperiodic nonlinear traveling waves of KdV) \cite{B89}. Specifically, the double cnoidal wave can be shown to be the proper spatially periodic generalization of the two-soliton solution of KdV \cite{HB91}. 

Back to our example, Fig.~\ref{fig:a2a4}(b) shows the evolution of the phase velocities $v_p(k)$ (in the simulations) of modes $k=2$ and $4$. The oscillations are caused by energy exchanges (interaction) between even modes. A low-pass filter is applied to evaluate a time-averaged phase velocity for mode 2, shown as the black curve. It is surprising to see that while $|\eta_2|$, which is the amplitude of Fourier mode 2, is growing slowly, its phase velocity maintains at $v_p(k=2)\approx -49.05$, which is independent of the phase velocity of mode 4, $v_p(k=4)\approx -127$.

The rapid transition during $t\in[1.1,1.25]$ is characterized by a change of propagation direction in the physical domain. At the same time, in the Fourier domain, modes 2 and 4 become comparable in energy content. Visually, this observation is similar to soliton collisions: when the peak of mode 2 is ``caught'' by that of mode 4, an elevation of the profile is observed. Subsequently, a depression of the profile is seen in Fig.~\ref{fig:a2a4}(c) for $t\in[1.147,1.181]$ as the waves separate. However, while soliton collision (in the sense of Zabusky and Kruskal \cite{ZK65}) leave the two interacting waves' profiles and propagation velocities unchanged upon collision, the interaction of the nonlinear periodic waves just described in current study results in the waves ultimately separating into what look like two localized solitons. The left-propagating components in the Fourier decomposition dramatically decrease in energy, and the profile appears as a standing wave at $t\approx 1.181$. Subsequently, all Fourier modes in the system merge into a right-propagating profile with phase velocity $v_p\approx 341.7$, which no longer follows the linear prediction from the dispersion relation, thus highlighting the strongly nonlinear interaction that has just occurred.

This result could also have been anticipated from Fig.~\ref{fig:a2a4}(a) wherein, after the transition, the localized solitons are seen to contain a wider energy spectrum than that of the period-four traveling wave (from before the transition). Therefore, the linear phase velocity of the leading mode is no longer predictive of the nonlinear wave speed. On the other hand, according to the parameters used for the example shown in Fig.~\ref{fig:a2a4}, the relatively large values of $a_1$, $a_2$ and $a_3$ indicate that the KdV-terms dominate in Eq.~\eqref{eq:KdV-KSV}. The interaction between Fourier modes becomes more and more intense, as the nonlinearity becomes more and more important, which results in a large wave amplitude in the physical domain. As discussed in \cite{YC21b}, this transition is indicative of (and can be explained by) the spectral instability \cite{KP13} of the period-four nonlinear traveling wave. Eventually, the wave evolves into the two localized solitons observed, which are visually similar to the $\sech^2$ solutions of KdV. In this case, the determination of the propagation velocity would involve all of the system parameters, as for the $\sech^2$ solution of KdV \cite{ZK65}.

Finally, it is worth comparing and contrasting the dynamics just described with those discussed in \cite{YC21b} under the long-wave equation~\eqref{eq:ferro_KS} for a ferrofluid thin film. Before the transition, the long-lived multiperiodic waves discussed herein and in \cite{YC21b} are qualitatively similar. However, after the transition, the period-four nonlinear traveling wave in~\cite{YC21b} merges with a period-two traveling wave, and the resultant propagation velocity is still well predicted by the linear theory. Under the KdV-KSV equation~\eqref{eq:KdV-KSV}, however, the transition is more intricate than the examples in \cite{YC21b} under Eq.~\eqref{eq:ferro_KS}: the wave profile changes from a period-four nonlinear traveling wave into what appear to be two localized soliton-like shapes. The mechanism of this transition, which to best of our knowledge has not been reported before, is still under investigation. Nevertheless, 25 years after \cite{CV95}, our numerical results shed light on the intriguing nonlinear dynamics of nonlinear periodic waves as dissipative solitons. 

\section{Conclusion}

Motivated by our ability to ``tune'' a spatially nonuniform magnetic field to turn a linearly unstable circular ferrofluid interface (confined in a Hele-Shaw cell) into a spinning ``gear'' \cite{YC21}, we previously derived a model long-wave equation~\eqref{eq:ferro_KS} for driven ferrofluids \cite{YC21b}. When the spinning droplet interface is ``unwrapped'' onto $x\in[0,2\pi]$, it is a manifestation of a nonlinear periodic traveling wave solution of a nonconservative long-wave equation. This appears to be a novel finding in the context of dissipative solitons. 

In this short note, we returned to the classical model equation~\eqref{eq:KdV-KSV} exhibiting dissipative solitons \cite{CV95}, and we demonstrated that it also features sustained nonlinear periodic traveling wave solutions. Despite the advective nonlinearities in Eq.~\eqref{eq:KdV-KSV}  (from \cite{CV95}) and Eq.~\eqref{eq:ferro_KS} (from \cite{YC21b}) being substantially different due to the different physics, the nonlinear periodic waves as dissipative solitons that emerge were shown to have similar features. For example, these solutions (despite being nonlinear waves) are characterized by low-dimensional Fourier decompositions, thus their propagation velocity can be well predicted by the phase velocity (calculated from the linear dispersion relation) of the leading Fourier mode. Another point of commonality is that the linear instability of the flat base state, which gives rise to the nonlinear periodic waves, is controlled by the ratio of the coefficients of the second and fourth order (anti-)diffusive terms in the respective PDEs. Thus, `control' of the period of the traveling wave solutions is possible. The transition between different nonlinear periodic states occurs when a spectrally unstable nonlinear traveling wave is subjected to global perturbations of with a specific wavenumber.

Importantly, model equations such as \eqref{eq:KdV-KSV} and \eqref{eq:ferro_KS} are \emph{nonintegrable} models in which one can observe generalizations of the so-called polycnoidal waves (multiperiodic nonlinear traveling waves) \cite{B89}. While the double cnoidal wave problem for integrable models like KdV is well understood \cite{HB91}, there is no equivalent theoretical understanding for nonintegrable models. Our recent work \cite{YC21b} on Eq.~\eqref{eq:ferro_KS} further suggests that the dominant balances for KdV found in \cite{HB91} may not be applicable to Eq.~\eqref{eq:ferro_KS}. Meanwhile, to the best of our knowledge, multiperiodic solutions to the classical Eq.~\eqref{eq:KdV-KSV} have not been studied prior to the present short note. In future work, it would be of interest to explore nonlinear periodic waves as dissipative solitons in other long-wave model equations for driven thin film flows (see, e.g., \cite{KRSV12}).

\begin{acknowledgments}
This research was supported by the US National Science Foundation under grant No.\ CMMI-2029540. 

\end{acknowledgments}

\bibliography{references}

\end{document}